\documentclass[aps,prb,reprint,amsmath,amssymb,nobibnotes]{revtex4-2}

\usepackage{bm}
\usepackage{graphicx}
\usepackage[utf8]{inputenc}
\usepackage{amsmath}
\usepackage{physics}
\usepackage{times}
\DeclareMathOperator{\arsinh}{arsinh}

\usepackage{hyperref} 
\hypersetup{colorlinks, allcolors=blue}

\begin{document}
\title{Stability of superconducting gap symmetries arising from antiferromagnetic magnons}

\author{Chi Sun}
\affiliation{\mbox{Center for Quantum Spintronics, Department of Physics, Norwegian University of Science and Technology, NO-7491 Trondheim, Norway}}
\author{Kristian M{\ae}land}
\affiliation{\mbox{Center for Quantum Spintronics, Department of Physics, Norwegian University of Science and Technology, NO-7491 Trondheim, Norway}}
\author{Asle Sudb{\o}}
\email[Corresponding author: ]{asle.sudbo@ntnu.no}
\affiliation{\mbox{Center for Quantum Spintronics, Department of Physics, Norwegian University of Science and Technology, NO-7491 Trondheim, Norway}}


\begin{abstract}
We consider a planar heterostructure consisting of a normal metal in proximity to an antiferromagnetic insulator, with an interlayer exchange coupling between the metal and the insulator. The coupling to the two sublattices of the antiferromagnetic insulator is allowed to be asymmetric. An effective electron-electron interaction in the normal metal, mediated by antiferromagnetic magnons in the insulator, is derived to second order in the interlayer exchange coupling.
Particular emphasis is placed on including analytically derived expressions for the effective interactions including Umklapp processes in the solutions to the superconducting gap equation. The gap equation is first solved at the critical temperature as an eigenvalue problem by linearizing the gap equation. The eigenvectors yield information on the symmetry of the superconducting gap at the onset of superconductivity, and we derive a phase diagram for the order parameter in this case. In the various regimes of the phase diagram, we find $p$-wave, $f$-wave, and $d$-wave superconductivity, with $p$-wave superconductivity in the dominant part of the phase diagram. Umklapp processes, that come into play with increasing size of the Fermi surface, yield $f$- and $d$-wave symmetries as the preferred symmetries when band filling approaches half filling. To investigate the stability of this order parameter symmetry as the temperature is lowered, we also consider the nonlinear gap equation at zero temperature. We conclude that the phase diagram and the symmetries of the superconducting order parameter essentially are left intact as the temperature is lowered to zero temperature.     
\end{abstract}

\maketitle

\section{Introduction}
The interaction between a magnetic insulator (MI) and an adjacent nonmagnetic material forms a major research topic in the field of spintronics. Spin-orbit torque \cite{shao2021roadmap,ramaswamy2018recent,gambardella2011current} and spin pumping \cite{heinrich2011spin,cheng2014spin} have been intensively studied to realize interconversion between electronic spin current in the nonmagnetic normal metal (NM) and magnonic spin current in the MI. Based on this, versatile electrically manipulated devices \cite{shao2021roadmap} such as magnetic recording units \cite{avci2017current}, nano-oscillators \cite{chen2016spin,collet2016generation,evelt2018spin} and domain-wall racetracks \cite{velez2019high} are developed. 
In addition, it has been proposed that superconductivity can be introduced in  MI/NM heterostructures \cite{rohling2018superconductivity,fjaerbu2019superconductivity,erlandsen2019enhancement, thingstad2021eliashberg, maeland2022topological, brekke2023interfacial, ExpMagnonInducedHeterostruct}. 
Compared to conventional phonon-mediated superconductivity, the magnons in the MI are responsible for mediating an attractive interaction between the electrons in the neighboring NM, leading to the formation of a superconducting state. The magnon-mediated superconductivity has the potential to be enhanced by engineering the properties of the MI such as magnetic anisotropy \cite{johansen2019magnon} and magnon gap \cite{fjaerbu2019superconductivity}. In addition, the MI/NM interface also plays a crucial role in manipulating the resulting superconductivity \cite{erlandsen2019enhancement,thingstad2021eliashberg, maeland2022topological, brekke2023interfacial}. 

The ferromagnetic insulator (FMI) combines the advantages of ferromagnetism and insulators, which is highly needed for developing low dissipation spintronic devices \cite{serga2010yig,Meng2018Mar}. Magnon-mediated superconductivity has been proposed in a FMI/NM/FMI trilayer \cite{rohling2018superconductivity}, in which a triplet $p$-wave superconducting pairing with critical temperature in the interval between 1 and 10~K is found. On the other hand, emerging research interests have recently been focused on antiferromagnetic insulators (AFMIs) with compensated magnetic moments, which possess a higher degree of stability and lower power consumption compared with FMIs \cite{jungwirth2016antiferromagnetic,sun2022enhanced}.
Similarly, it has been shown theoretically that an electron-electron interaction yielding superconductivity can be mediated by antiferromagnetic magnons in an AFMI/NM bilayer \cite{erlandsen2019enhancement} and an AFMI/NM/AFMI trilayer \cite{fjaerbu2019superconductivity,thingstad2021eliashberg}. In Refs.~\cite{erlandsen2019enhancement, thingstad2021eliashberg} it is shown that asymmetric coupling to the sublattices of the AFMI gives rise to a potential enhancement of the critical temperature due to squeezing of magnons \cite{Kamra2019antiferromagnetic}.

\begin{figure}[ht]
\centering
\includegraphics[width=0.95\columnwidth]{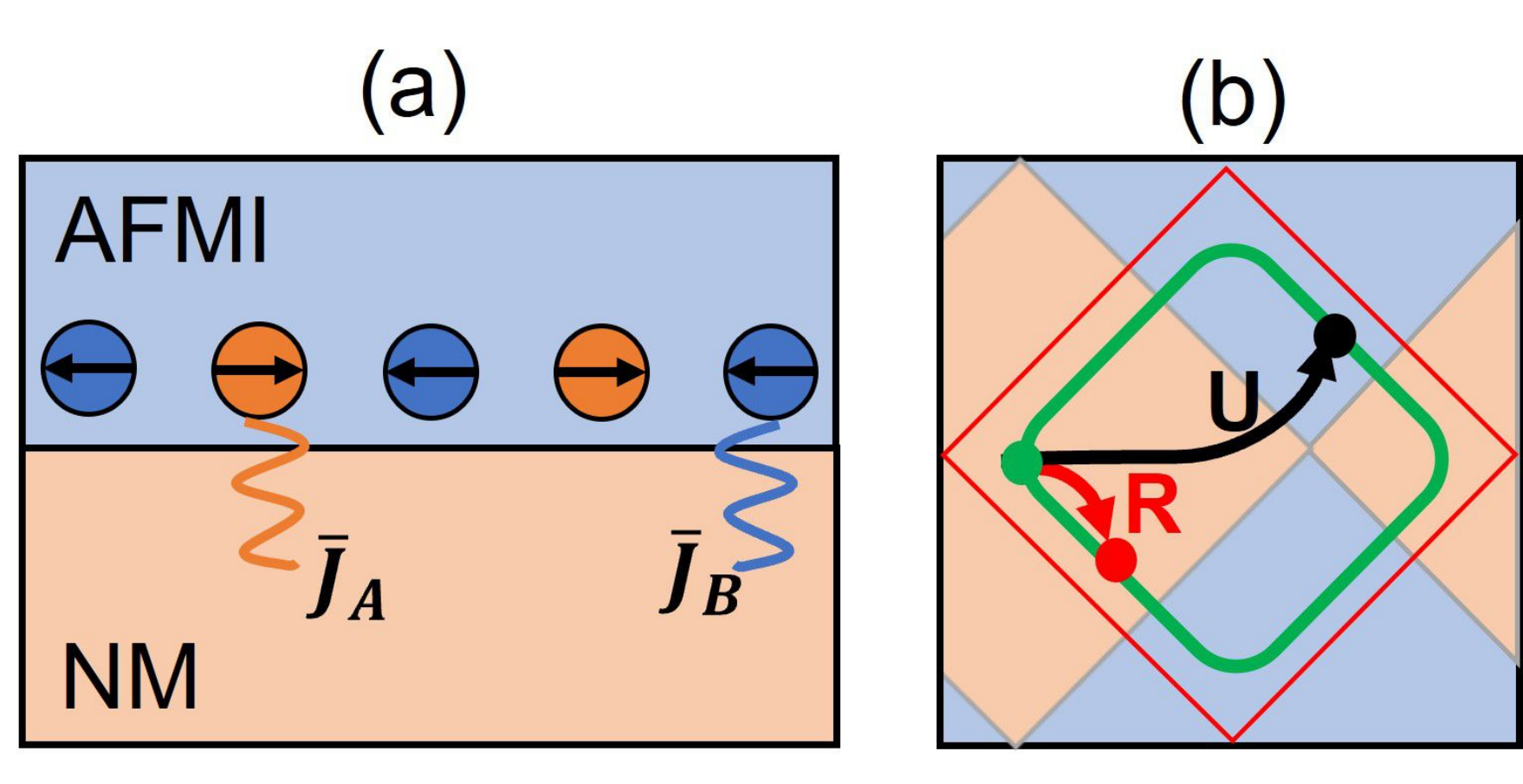}
  \caption{(Color online) (a) Schematic representation of the bilayer structure considered in this paper, with a normal metal (NM) and an antiferromagnetic insulator (AFMI) on a bipartite lattice (orange and blue). The spins of the AFMI on the two sublattices interact with the spins of the electrons with exchange coupling constants that are allowed to be different on each sublattice, but are uniform within each sublattice. (b) Illustration of a regular (R) and Umklapp (U) scattering process. The entire square is the Brillouin zone for the electrons, while the red square illustrates the reduced magnetic Brillouin zone (RBZ). An example Fermi surface is shown in green. In a regular scattering process, magnons can transfer momenta $\boldsymbol{q}$ between electrons in the range defined by the RBZ, i.e., within the orange region. Umklapp processes are necessary to reach momenta outside the RBZ, i.e., in the blue region.   
  }%
  \label{fig:structure}
\end{figure}

In the AFMI, magnons reside in a reduced Brillouin zone (RBZ) compared to the full first Brillouin zone (1BZ) for electrons in the NM. This introduces electron-magnon scattering of two types, i.e., regular and Umklapp [see Fig.~\ref{fig:structure}(b)]. The electrons are scattered with a momentum within the magnon RBZ through regular scattering. This is the only relevant process for a small Fermi surface (FS), like those considered in Ref.~\cite{erlandsen2019enhancement} for the AFMI/NM bilayer, where magnon-mediated $p$-wave superconductivity was found. In the Umklapp processes, the electrons are scattered out of the RBZ by receiving an additional momentum corresponding to a magnon lattice vector in the reciprocal space. This mechanism becomes important as the FS of the NM becomes larger and approaches half-filling, resulting in a $d$-wave phase based on the AFMI/NM/AFMI trilayer within a weak-coupling approach in Ref.~\cite{fjaerbu2019superconductivity} and within a strong-coupling approach in Ref.~\cite{thingstad2021eliashberg}. Ref.~\cite{thingstad2021eliashberg} also finds $f$-wave pairing close to half filling with sublattice coupling asymmetry.

In this paper, we study magnon-mediated superconductivity in an AFMI/NM bilayer [see Fig.~\ref{fig:structure}(a)] within a weak-coupling BCS framework, where both regular and Umklapp scatterings are treated on an equal footing. 
We analytically derive expressions for the effective electron-electron interactions mediated by magnons and numerically solve the resulting gap equation. Both the linearized gap equation close to the critical temperature and the nonlinear gap equation at zero temperature are solved. 
By modulating the asymmetry of interfacial sublattice exchange coupling and the chemical potential in the NM, transitions between different superconducting phases (i.e., $p$-, $d$- and $f$-waves) are achieved.

This paper is related to Refs.~\cite{fjaerbu2019superconductivity, erlandsen2019enhancement, thingstad2021eliashberg}. Refs.~\cite{ erlandsen2019enhancement, thingstad2021eliashberg} are restricted to temperatures close to the critical temperature of superconductivity, while Ref.~\cite{fjaerbu2019superconductivity} is limited to symmetric coupling to the AFMI sublattices and an approximate model with quadratic electron dispersion or circular FS is utilized. We use a tight-binding model to treat the noncircular FS when the chemical potential is tuned to approach half-filling. A study of the superconducting phase diagram at zero temperature including sublattice asymmetric coupling and an accurate account of the effect of Umklapp processes close to half-filling, is the main result of this paper. As such, the present paper improves on insights from previous works~\cite{ erlandsen2019enhancement, thingstad2021eliashberg} . 
Furthermore, our model provides both analytical and numerical insights into the AFMI magnon-mediated superconductivity, giving potential theoretical suggestions on how to realize and manipulate magnon-mediated superconductivity experimentally.

\section{Theory}
\subsection{Model}
The AFMI/NM bilayer system, illustrated in Fig.~\ref{fig:structure}(a), is modeled by the Hamiltonian 
\begin{equation}
    H = H_{\text{AFMI}} + H_{\text{NM}} + H_{\text{int}},
\end{equation}
which consists of the AFMI, NM and interfacial terms whose explicit expressions are given by
\begingroup
\allowdisplaybreaks
\begin{align}
    H_{\text{AFMI}} =& J\sum_{\langle i,j \rangle} \boldsymbol{S}_i \cdot \boldsymbol{S}_j - K\sum_{i}S_{iz}^2,\\
    H_{\text{NM}} =& -t\sum_{\langle i,j \rangle ,\sigma}c_{i\sigma}^{\dag} c_{j\sigma} - \mu \sum_{i,\sigma}c_{i\sigma}^\dag c_{i\sigma},\\
    H_{\text{int}} =& -2\bar{J}_A \sum_{i\in A}c_i^\dag \boldsymbol{\sigma} c_i \cdot \boldsymbol{S}_i -2\bar{J}_B \sum_{i\in B}c_i^\dag \boldsymbol{\sigma} c_i \cdot \boldsymbol{S}_i.
\end{align}
\endgroup
Here the sum over $\langle i,j \rangle$ includes all nearest neighboring sites. In the AFMI, $J$ and $K$ denote the antiferromagnetic exchange and easy-axis anisotropy (along $\hat{z}$), respectively. $\boldsymbol{S}_i$ is the spin operator at the lattice site $i$. In the NM, $t$ is the tight-binding hopping parameter and $\mu$ represents the chemical potential. $c_{i\sigma}^\dag$ $(c_{i\sigma})$ is the electron creation (annihilation) operator, which creates (annihilates) an electron with spin $\sigma$ at site $i$. At the AFMI/NM interface, $\bar{J}_A$ $(\bar{J}_B)$ parametrizes the sublattice-dependent interfacial exchange coupling. In the following, these interfacial strengths are described by the coupling asymmetry parameter $\Omega \equiv \bar{J}_A/\bar{J}_B$ with $\bar{J}_B \equiv \bar{J}$. The notation $c_{i} \equiv (c_{i\uparrow},c_{i\downarrow})^T$ is introduced and $\boldsymbol{\sigma}$ denotes the Pauli matrix vector in the spin space. In addition, we set $\hbar=a=1$, where $a$ is the lattice constant of the square lattice considered in this paper.

Applying the Holstein-Primakoff and Fourier transformations for the two sublattice spin operators in the AFMI, $H_{\text{AFMI}}$ is transferred into the momentum space in terms of the two individual sublattice magnons $a_{\boldsymbol{q}}$ and $b_{\boldsymbol{q}}$ (see Appendix \ref{app:HPandFT} for details). Next, the Bogoliubov transformation with $\alpha_{\boldsymbol{q}} = u_{\boldsymbol{q}} a_{\boldsymbol{q}} - v_{\boldsymbol{q}} b_{-{\boldsymbol{q}}}^\dag$ and $\beta_{\boldsymbol{q}} = u_{\boldsymbol{q}} b_{\boldsymbol{q}} - v_{\boldsymbol{q}} a_{-{\boldsymbol{q}}}^\dag$ is performed to diagonalize the AFMI Hamiltonian, in which $\alpha_{\boldsymbol{q}}$ and $\beta_{\boldsymbol{q}}$ are the eigenmagon operators defined as superpositions of $a_{\boldsymbol{q}}$ and $b_{\boldsymbol{q}}$. The resulting diagonalized Hamiltonian is given by
\begin{align}
    H_{\text{AFMI}} =& \sum_{{\boldsymbol{q}}\in \Diamond} \omega_{\boldsymbol{q}} (\alpha_{\boldsymbol{q}}^\dag \alpha_{\boldsymbol{q}} + \beta_{\boldsymbol{q}}^\dag \beta_{\boldsymbol{q}}),\\
    \omega_{\boldsymbol{q}} =& 2s\sqrt{(zJ+K)^2-z^2 J^2 \gamma_{\boldsymbol{q}}^2},
\end{align}
in which $\gamma_{\boldsymbol{q}} = (2/z) (\cos{q_x}+\cos{q_y})$ is the structure factor. $z$ and $s$ are the number of nearest neighbors and spin quantum number, respectively. $\Diamond $ denotes summation over the RBZ. The coherence factors in the Bogoliubov transformation are obtained as $u_{\boldsymbol{q}} = \cosh{\theta_{\boldsymbol{q}}}$ and $v_{\boldsymbol{q}} = \sinh{\theta_{\boldsymbol{q}}}$ with $\theta_{\boldsymbol{q}} = (1/2)\operatorname{artanh}[-Jz\gamma_{\boldsymbol{q}}/(zJ+K)]$.

The NM Hamiltonian can be diagonalized as 
\begin{align}
H_{\text{NM}}&=\sum_{\boldsymbol{k}\in \Box, \sigma} \epsilon_{\boldsymbol{k}} c_{\boldsymbol{k}\sigma}^{\dag}c_{\boldsymbol{k}\sigma},\\
\epsilon_{\boldsymbol{k}} &= -tz\gamma_{\boldsymbol{k}} - \mu,
\label{eq:FS}
\end{align}
in which $\Box$ denotes the sum over the full Brillouin zone. Here we utilized
\begin{align}
    c_{i\sigma} &= \frac{1}{\sqrt{N}}\sum_{\boldsymbol{k}\in \Box}c_{\boldsymbol{k}\sigma}e^{-i\boldsymbol{k}\cdot \boldsymbol{r}_i} \notag\\&= \frac{1}{\sqrt{N}}\sum_{\boldsymbol{k}\in \Diamond}(c_{\boldsymbol{k}\sigma}e^{-i\boldsymbol{k}\cdot \boldsymbol{r}_i} + c_{\boldsymbol{k}+\boldsymbol{G},\sigma}e^{-i(\boldsymbol{k}+\boldsymbol{G})\cdot \boldsymbol{r}_i})
\label{eq:Fourier_NM}
\end{align}
in the Fourier transformation,
where $N$ is the number of lattice sites at the AFMI/NM interface. $\boldsymbol{G} \equiv \pi(\hat{x}+\hat{y})$ is the reciprocal lattice vector, which occurs in the Umklapp scattering processes.

Utilizing the transformations shown in the Appendix \ref{app:HPandFT} and Eq.~(\ref{eq:Fourier_NM}), the interfacial Hamiltonian $H_{\text{int}}=H_{\text{int}}^{(A)}+H_{\text{int}}^{(B)}$ is written in terms of diagonalized operators as
\begin{widetext}
\begin{align}
 H_{\text{int}}^{(A)} =& \Omega V \sum_{\boldsymbol{k}\in \Box,\boldsymbol{q} \in \Diamond}[ (u_{\boldsymbol{q}}\alpha_{\boldsymbol{q}} + v_{\boldsymbol{q}}\beta_{-\boldsymbol{q}}^\dag)(c_{\boldsymbol{k} + \boldsymbol{q},\downarrow}^\dag c_{\boldsymbol{k}\uparrow} + c_{\boldsymbol{k} + \boldsymbol{q} + \boldsymbol{G},\downarrow}^\dag c_{\boldsymbol{k}\uparrow} ) + (u_{\boldsymbol{q}}\alpha_{-\boldsymbol{q}}^\dag + v_{\boldsymbol{q}}\beta_{\boldsymbol{q}})(c_{\boldsymbol{k} + \boldsymbol{q},\uparrow}^\dag c_{\boldsymbol{k}\downarrow} + c_{\boldsymbol{k} + \boldsymbol{q} + \boldsymbol{G},\uparrow}^\dag c_{\boldsymbol{k}\downarrow})], \label{eq:HintA}\\
 H_{\text{int}}^{(B)} =& V \sum_{\boldsymbol{k}\in \Box,\boldsymbol{q} \in \Diamond}[ (u_{\boldsymbol{q}}\beta_{\boldsymbol{q}} + v_{\boldsymbol{q}}\alpha_{-\boldsymbol{q}}^\dag)(c_{\boldsymbol{k} + \boldsymbol{q},\uparrow}^\dag c_{\boldsymbol{k}\downarrow} - c_{\boldsymbol{k} + \boldsymbol{q} + \boldsymbol{G},\uparrow}^\dag c_{\boldsymbol{k}\downarrow} ) + (u_{\boldsymbol{q}}\beta_{-\boldsymbol{q}}^\dag + v_{\boldsymbol{q}}\alpha_{\boldsymbol{q}})(c_{\boldsymbol{k} + \boldsymbol{q},\downarrow}^\dag c_{\boldsymbol{k}\uparrow} - c_{\boldsymbol{k} + \boldsymbol{q} + \boldsymbol{G},\downarrow}^\dag c_{\boldsymbol{k}\uparrow})], \label{eq:HintB}
\end{align}
\end{widetext}
which couples the electrons in the NM with the A and B sublattice magnons in the AFMI. Here we define $V \equiv -2\bar{J}\sqrt{s}/\sqrt{N}$. 

\subsection{Effective interaction}
The full Hamiltonian of the AFMI/NM bilayer system can be written as $H = H_0 + \eta H_1$, in which we define
\begin{align}
    H_0 &\equiv H_\text{AFMI} + H_\text{NM} \notag\\&= \sum_{{\boldsymbol{q}}} \omega_{\boldsymbol{q}} (\alpha_{\boldsymbol{q}}^\dag \alpha_{\boldsymbol{q}} + \beta_{\boldsymbol{q}}^\dag \beta_{\boldsymbol{q}}) + \sum_{\boldsymbol{k} \sigma} \epsilon_{\boldsymbol{k}} c_{\boldsymbol{k}\sigma}^{\dag}c_{\boldsymbol{k}\sigma},\\
    \eta H_1 &= \eta H_1^{(A)} + \eta H_1^{(B)} \equiv H_{\text{int}}^{(A)} + H_{\text{int}}^{(B)}. 
\end{align}
In order to obtain an effective electron-electron interaction $H_\text{pair}$ mediated by virtual magnons, we perform a canonical transformation by treating $\eta H_1$ as a perturbation. $\eta$ is a dummy-variable to count powers in the perturbation expansion. The resulting effective Hamiltonian takes the form of $H_\text{eff} = H_0 + H_\text{pair}$ with (see Appendix \ref{app:HPair} for details)
\begin{equation}
\label{eq:HpairMain}
    H_{\text{pair}} = \sum_{\boldsymbol{k}\boldsymbol{k}^{'}} V_{\boldsymbol{k}\boldsymbol{k}^{'}} c_{\boldsymbol{k}\uparrow}^\dag c_{-\boldsymbol{k}\downarrow}^\dag c_{-\boldsymbol{k}^{'}\downarrow} c_{\boldsymbol{k}^{'}\uparrow},
\end{equation}
in which
\begin{align}
V_{\boldsymbol{k}\boldsymbol{k}^{'}} =& -V^2 \frac{2\omega_{\boldsymbol{q}}}{(\epsilon_{\boldsymbol{k}^{'}}-\epsilon_{\boldsymbol{k}})^2 - \omega_{\boldsymbol{q}}^2} A(\boldsymbol{q}, \Omega), \label{eq:Veff} \\
A(\boldsymbol{q}, \Omega) =& \frac{1}{2} (\Omega^2 + 1) (u_{\boldsymbol{q}}^2 + v_{\boldsymbol{q}}^2) + 2\Theta_{\boldsymbol{q}}\Omega u_{\boldsymbol{q}}v_{\boldsymbol{q}},\\
\boldsymbol{q} =& \begin{cases} \boldsymbol{k} + \boldsymbol{k}^{'}, & \boldsymbol{k} + \boldsymbol{k}^{'} \in \Diamond \\
\boldsymbol{k} + \boldsymbol{k}^{'} + \boldsymbol{G},& \boldsymbol{k} + \boldsymbol{k}^{'} \notin \Diamond
\end{cases},\\
\Theta_{\boldsymbol{q}}=&\begin{cases} 1, & \boldsymbol{k} + \boldsymbol{k}^{'} \in \Diamond \\
-1, & \boldsymbol{k} + \boldsymbol{k}^{'} \notin \Diamond
\end{cases}.
\end{align}
The effective interaction $V_{\boldsymbol{k}\boldsymbol{k}^{'}}$ in Eq. (\ref{eq:Veff}) is of the standard form well-known for electron-phonon interactions, apart from the factor 
$A(\boldsymbol{q}, \Omega)$. This is a factor that boosts the strength of the magnon-mediated electron-electron interaction. The boosting has two origins: i) Varying $\Omega$ from $1$ to $0$ changes the combination of coherence factors in the interaction from $(u_{\boldsymbol{q}} + v_{\boldsymbol{q}})^2$ to $u^2_{\boldsymbol{q}} + v^2_{\boldsymbol{q}}$. Since 
$u_{\boldsymbol{q}}$ and $v_{\boldsymbol{q}}$
have opposite signs and tend to cancel for small $\boldsymbol{q}$, this vastly enhances the strength of the interaction for small-momenta scattering \cite{erlandsen2019enhancement, erlandsen2020magnon}. ii) Umklapp scattering implies that the factor $\Theta_{\boldsymbol{q}}$ takes on the opposite sign compared to regular scattering. Thus, even for $\Omega=1$ one obtains the combination $(u_{\boldsymbol{q}} - v_{\boldsymbol{q}})^2$ instead of $(u_{\boldsymbol{q}} + v_{\boldsymbol{q}})^2$, again leading to a strengthening of the interaction \cite{fjaerbu2019superconductivity,erlandsen2019enhancement}. 

\section{RESULTS AND DISCUSSION}
Consider the odd part of the pairing potential $V^{O(\boldsymbol{k})}_{\boldsymbol{k}\boldsymbol{k^{'}}}=\frac{1}{2}(V_{\boldsymbol{k}\boldsymbol{k}^{'}}-V_{-\boldsymbol{k},\boldsymbol{k}^{'}})$ for the $S_z=0$ spin triplet channel, i.e., the typical condensation channel for magnon-mediated superconductivity. The BCS gap function is defined as $\Delta_{\boldsymbol{k}}=-\sum_{\boldsymbol{k}^{'}}V^{O(\boldsymbol{k})}_{\boldsymbol{k}\boldsymbol{k}^{'}}\langle c_{-\boldsymbol{k}^{'}\uparrow}c_{\boldsymbol{k}^{'}\downarrow} + c_{-\boldsymbol{k}^{'}\downarrow}c_{\boldsymbol{k}^{'}\uparrow} \rangle/2$. Within the standard weak-coupling mean-field theory approach \cite{erlandsen2019enhancement, sigrist1991phenomenological}, the gap function becomes
\begin{equation}
    \Delta_{\boldsymbol{k}}=-\sum_{\boldsymbol{k}^{'}}V^{O(\boldsymbol{k})}_{\boldsymbol{k}\boldsymbol{k}^{'}}\frac{\Delta_{\boldsymbol{k}^{'}}}{2E_{\boldsymbol{k}^{'}}}\tanh\pqty{\frac{E_{\boldsymbol{k}^{'}}}{2k_B T}},
\label{eq:gap}
\end{equation}
which is a nonlinear equation with respect to $\Delta_{\boldsymbol{k}}$ with  $E_{\boldsymbol{k}}=\sqrt{\epsilon_{\boldsymbol{k}}^2+|\Delta_{\boldsymbol{k}}|^2}$. On the other hand, the even part of the pairing potential $V^{E(\boldsymbol{k})}_{\boldsymbol{k}\boldsymbol{k^{'}}}=\frac{1}{2}(V_{\boldsymbol{k}\boldsymbol{k}^{'}}+V_{-\boldsymbol{k},\boldsymbol{k}^{'}})$ corresponds to the spin singlet channel, where $\Delta_{\boldsymbol{k}}=-\sum_{\boldsymbol{k}^{'}}V^{E(\boldsymbol{k})}_{\boldsymbol{k}\boldsymbol{k}^{'}}\langle c_{-\boldsymbol{k}^{'}\uparrow}c_{\boldsymbol{k}^{'}\downarrow} - c_{-\boldsymbol{k}^{'}\downarrow}c_{\boldsymbol{k}^{'}\uparrow} \rangle/2$. Its gap equation takes the same form as Eq.~(\ref{eq:gap}), except that $V^{O(\boldsymbol{k})}_{\boldsymbol{k}\boldsymbol{k^{'}}}$ is replaced by $V^{E(\boldsymbol{k})}_{\boldsymbol{k}\boldsymbol{k^{'}}}$.

\subsection{Gap equation at critical temperature}
When the temperature $T$ approaches its critical value $T_c$ from below, the magnitude of $\Delta_{\boldsymbol{k}}$ in $E_{\boldsymbol{k}}=\sqrt{\epsilon_{\boldsymbol{k}}^2+|\Delta_{\boldsymbol{k}}|^2}$ can be treated as negligible, giving rise to a linearized  gap equation
\begin{equation}
    \Delta_{\boldsymbol{k}}=-\sum_{\boldsymbol{k}^{'}}V^{O(\boldsymbol{k})}_{\boldsymbol{k}\boldsymbol{k}^{'}}\frac{\Delta_{\boldsymbol{k}^{'}}}{2|\epsilon_{\boldsymbol{k}^{'}}|}\tanh(\frac{|\epsilon_{\boldsymbol{k}^{'}}|}{2k_BT_c}).
\label{eq:gap_linear}
\end{equation}
In the following, we assume that the gap is nonzero only close to the FS for momenta such that $|\epsilon_{\boldsymbol{k}}|, |\epsilon_{\boldsymbol{k}^{'}}|<\omega_c$, with $\omega_c = 2szJ$ the magnon cutoff energy at the Brillouin zone boundary. Furthermore, $\tanh(|\epsilon_{\boldsymbol{k}^{'}}|/2k_BT_c)/2|\epsilon_{\boldsymbol{k}^{'}}|$ is peaked at the FS, justifying a FS average \cite{sigrist1991phenomenological},
\begin{equation}
    \Delta_{\boldsymbol{k}} = -D(\mu)\langle V^{O(\boldsymbol{k})}_{\boldsymbol{k}\boldsymbol{k^{'}}} \Delta_{\boldsymbol{k}^{'}} \rangle_{\boldsymbol{k}^{'},\text{FS}}\int_{-\omega_c}^{\omega_c}d\epsilon \frac{1}{2|\epsilon|}\tanh(\frac{|\epsilon|}{2k_BT_c}),
\end{equation}
where $D(\mu)$ is the density of states per spin on the FS (see Appendix \ref{app:DOS} for its explicit expression) and $\langle\ \rangle_{\boldsymbol{k}^{'},\text{FS}}$ denotes the angular average over the FS. Next, a dimensionless coupling constant is defined as \cite{fossheim2004superconductivity} 
\begin{equation}
   \frac{1}{\lambda} = \int_{-\omega_c}^{\omega_c}d\epsilon \frac{1}{2|\epsilon|}\tanh(\frac{|\epsilon|}{2k_BT_c}) \approx \ln(\frac{1.13\omega_c}{k_B T_c}),
\label{eq:1/lamb}
\end{equation}
in which the weak-coupling limit ($\lambda \ll 1$) is assumed. In terms of $\lambda$, the gap function can be written as 
\begin{equation}
    \lambda \Delta_{\boldsymbol{k}} = -D(\mu)\langle V^{O(\boldsymbol{k})}_{\boldsymbol{k}\boldsymbol{k^{'}}} \Delta_{\boldsymbol{k}^{'}} \rangle_{\boldsymbol{k}^{'},\text{FS}},
\end{equation}
which becomes an eigenvalue problem by picking discrete points equidistantly placed on the FS for $\boldsymbol{k}$ and $\boldsymbol{k}^{'}$ to construct $V^{O(\boldsymbol{k})}_{\boldsymbol{k}\boldsymbol{k^{'}}}$ as a matrix. Consequently, $\lambda$ and $\Delta_{\boldsymbol{k}}$ can be solved as eigenvalues and eigenvectors, respectively. Here we utilize the largest eigenvalue for $\lambda$ to determine $T_c$ and its corresponding eigenvector to get the gap structure information. While this procedure determines $T_c$, it only determines the gap-amplitude  up to a multiplicative constant. We will return to this below when we consider solutions to the gap equation for $T < T_c$. Based on Eq.~(\ref{eq:1/lamb}), the estimated critical temperature $T_c$ is given by
\begin{equation}
    k_B T_c = 
    \frac{2 e^\gamma}{\pi} \omega_c e^{-1/\lambda}
    \approx 1.13 \omega_c e^{-1/\lambda},
\end{equation}
where $\gamma =0.577\dots$ is Euler's constant.
\begin{figure*}[ht]
  \centering
  \includegraphics[width=0.9\linewidth]{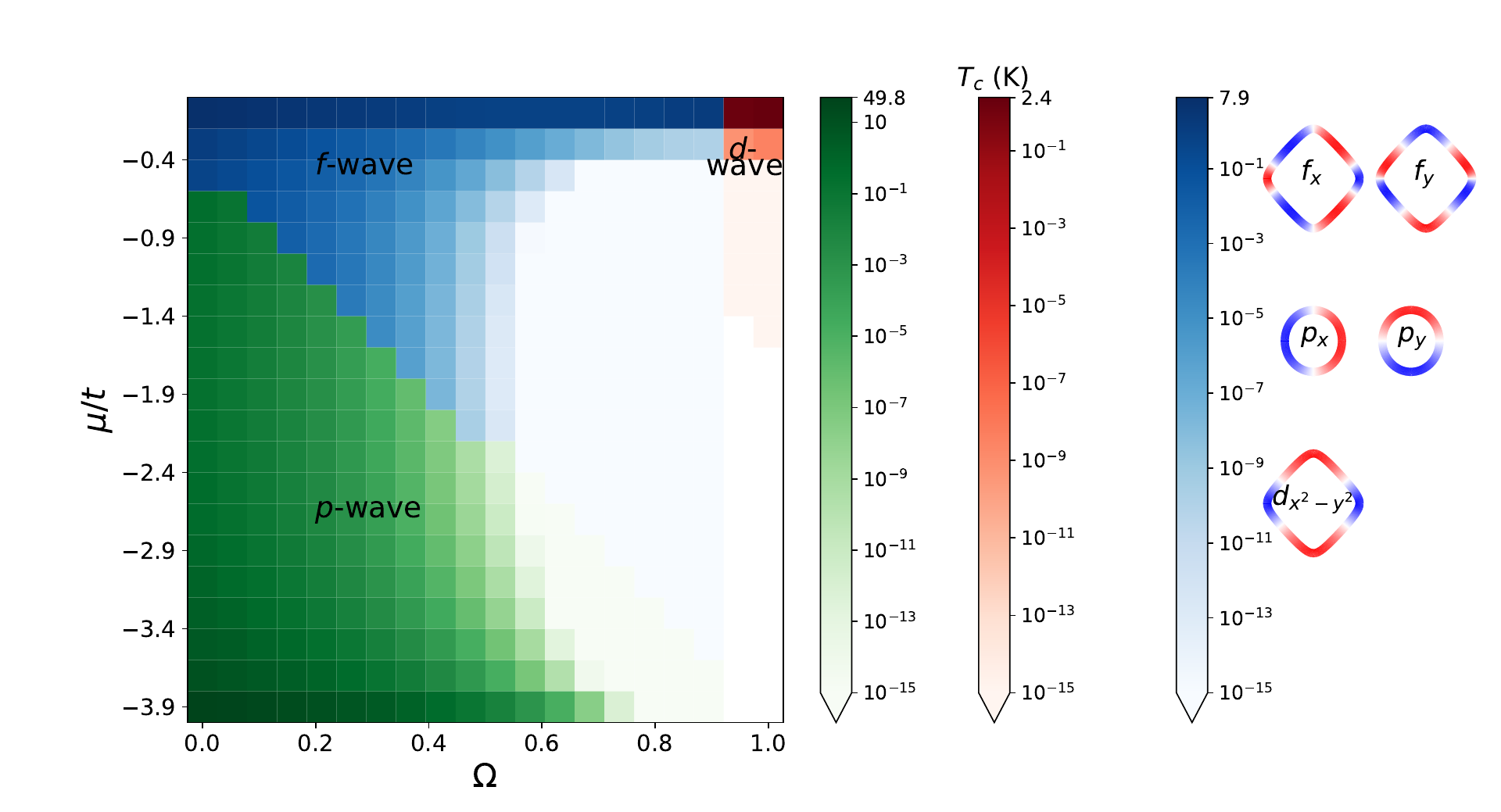}
  \caption{(Color online) Phase diagram of the critical temperature $T_c$ in terms of sublattice coupling asymmetry $\Omega$ and chemical potential $\mu$. The green, blue and red colors correspond to spin-triplet $p$-wave, spin-triplet $f$-wave, and spin singlet $d$-wave phases, respectively. The corresponding gap structures are shown on the right, red for positive and blue for negative $\Delta_{\boldsymbol{k}}$ on the Fermi surface. Here, we have used parameter values $J=5$ meV, $K=J/2000$, $\bar{J}=20$ meV, $s=1$, $z=4$ and $t=1$ eV.
  }%
  \label{fig:Tc}
\end{figure*}
As described above, we follow the same procedure to obtain $T_c$ for the spin singlet channel with $V^{E(\boldsymbol{k})}_{\boldsymbol{k}\boldsymbol{k^{'}}}$ and then determine the final $T_c$ as the greatest value obtained from these two channels. Employing plausible material parameters obtained from experiments \cite{fjaerbu2019superconductivity, erlandsen2019enhancement, Samuelsen1970inelastic, Kajiwara2010transmission}, the critical temperature $T_c$ is plotted as a phase diagram in terms of  the chemical potential $\mu$ and the asymmetry parameter $\Omega \equiv \bar{J}_A/\bar{J}_B$ in Fig.~\ref{fig:Tc}, where different colors are utilized to show the symmetries of their corresponding gaps (i.e., eigenvectors). In the green region with smaller $\mu$ and $\Omega$, we find spin triplet $p$-wave superconductivity. 
Since the square lattice is symmetric under $\pi/2$ rotations, $p_x$ and $p_y$ are degenerate solutions of the eigenvalue problem. Hence, all linear combinations of $p_x$ and $p_y$ are solutions, and we refer to this as $p$-wave. Possible solutions include a time-reversal-symmetry-broken $p_x + ip_y$ gap \cite{sigrist1991phenomenological}. Our definitions of gap symmetries are illustrated in Fig.~\ref{fig:Tc}.

In the $p$-wave region the highest $T_c \sim 50$ K is achieved at the largest asymmetry $\Omega=0$ and the lowest chemical potential $\mu = -3.9t$. 
At such low filling, the FS is small, ensuring that all scattering processes involve small magnitudes of magnon momenta $\boldsymbol{q}$ where the strongest coupling is expected for small $\Omega$ \cite{erlandsen2019enhancement, erlandsen2020magnon}. A similar increase in $T_c$ for small FSs and small $\Omega$ was found within a strong-coupling calculation in Ref.~\cite{thingstad2021eliashberg}.

As $\Omega < 1$ decreases, the electron-magnon coupling with small momentum transfer increases due to squeezing of antiferromagnetic magnons \cite{Kamra2019antiferromagnetic}. We get $V_{\boldsymbol{k}\boldsymbol{k}'}^{O(\boldsymbol{k})} < 0$ with large magnitude for $\boldsymbol{k} \approx \boldsymbol{k}'$ and $V_{\boldsymbol{k}\boldsymbol{k}'}^{O(\boldsymbol{k})} > 0$ with large magnitude for $\boldsymbol{k} \approx -\boldsymbol{k}'$ promoting $p$-wave spin triplet gaps. On the other hand, $V_{\boldsymbol{k}\boldsymbol{k}'}^{E(\boldsymbol{k})} > 0$ with increasing magnitude when $\boldsymbol{k} \approx \boldsymbol{k}'$ which prevents solutions to the gap equation for spin singlet gaps. As a result spin triplet superconductivity is preferred. 

When the size of the FS increases, Umklapp processes become relevant as two points on the FS may be further apart than the extent of the RBZ. A consequence of this is that $V_{\boldsymbol{k}\boldsymbol{k}'}^{O(\boldsymbol{k})}$, viewed as a function of $\boldsymbol{k}'$ on the FS with $\boldsymbol{k}$ fixed to a point of the FS, develops four more sign changes. Hence, it eventually becomes preferable for the gap to have six sign changes, $f$-wave instead of $p$-wave. The strong peaks in $V_{\boldsymbol{k}\boldsymbol{k}'}^{O(\boldsymbol{k})}$ for $\boldsymbol{k} \approx \pm \boldsymbol{k}'$ increase in magnitude for decreasing $\Omega$. Hence, more Umklapp processes are needed to make an $f$-wave gap preferred. This explains why the transition from $p$-wave to $f$-wave occurs at larger $\mu$ for smaller $\Omega$. Similar to the $p$-wave case, there is a degeneracy of $f_x$ and $f_y$. Any linear combination of $f_x$ and $f_y$ solves the linearized gap equation, and we refer to this as $f$-wave.

For $\Omega = 1$ there is no enhancement of the coupling for $\boldsymbol{k} \approx \pm \boldsymbol{k}'$. In fact, $V_{\boldsymbol{k}\boldsymbol{k}'}^{O(\boldsymbol{k})} > 0$ when $\boldsymbol{k} \approx \boldsymbol{k}'$ preventing solutions to the gap equation for spin triplet gaps. $V_{\boldsymbol{k}\boldsymbol{k}'}^{E(\boldsymbol{k})}$ is positive (but anisotropic) for all $\boldsymbol{k},\boldsymbol{k}'$ on the FS. Hence the spin singlet gap prefers a sign changing $d$-wave over $s$-wave. So, close to $\Omega = 1$, spin singlet $d$-wave is preferred, which results in the red region in Fig.~\ref{fig:Tc}. Unlike the spin triplet case, the spin singlet eigenvalue problem is not degenerate, and we find that a $d_{x^2-y^2}$ gap symmetry is preferred. Note that $T_c$ increases rapidly when approaching half filling, where the Umklapp enhancement of the interaction is most active. Further away from half filling the coupling is too weak to yield a measurable $T_c$ in any pairing channel.

Similar phase transitions can also be achieved based on the strong-coupling Eliashberg theory \cite{thingstad2021eliashberg}. This is an important point, since the boosting mechanism for $T_c$, involving both Umklapp and the coherence factors from diagonalizing the antiferromagnetic magnons, easily puts us in the strong-coupling regime. 
Ref.~\cite{thingstad2021eliashberg} finds an effective cutoff on the magnon spectrum that is lower than $\omega_c = 2szJ$, giving a reduction of the estimated $T_c$. 
The similarity of our present results to the strong-coupling results as far as the symmetry of the gap is concerned right at $T=T_c$, motivates us to also consider the low-temperature limit using a weak coupling treatment. This problem is considerably more demanding numerically, since the equations to be solved are non-linear. 

\subsection{Gap equation at zero temperature}
To investigate the low temperature behavior of the superconductor, we consider the gap equation at zero temperature. 
The main purpose of this is to investigate if the symmetry of the superconducting order parameter that we find at $T=T_c$ is altered in any significant way as the temperature is lowered. 
This is not an entirely trivial question. At $T=T_c$, as mentioned above, the gap-equation is linear. Hence, any linear combination of two solutions with the same eigenvalue  is also a solution. The situation is altered for $T < T_c$, where the gap-equation is non-linear. Consequently, one expects the degeneracy between solutions that exist at $T=T_c$ to be lifted. In addition, it is conceivable that solutions that are preferred at $T=T_c$ could be replaced by other solutions with different symmetries for $T<T_c$. Determining what happens requires an explicit calculation.    
To investigate the low--temperature regime, we consider the limit $T=0$, since we expect the temperature-dependence of the gap to saturate well above $T=0$. The present analysis therefore extends previous work using weak-coupling BCS-theory not including Umklapp, where only the $T=T_c$ case was studied in Ref.~\cite{erlandsen2019enhancement}. It also extends the work properly including Umklapp-processes \cite{thingstad2021eliashberg}, which only considered $T=T_c$.

\begin{figure*}[ht]
  \centering
  \includegraphics[width=0.9\linewidth]{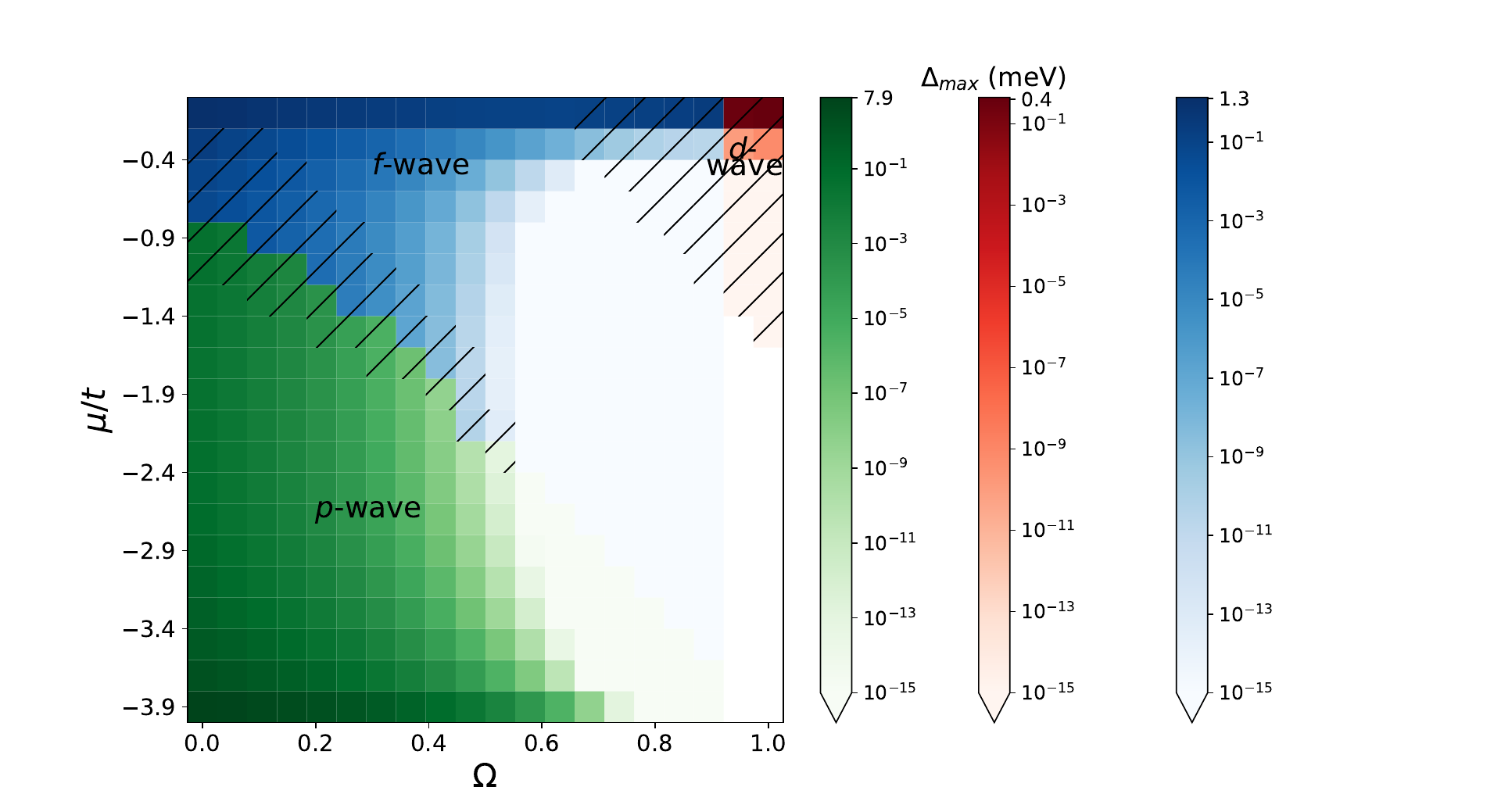}
  \caption{(Color online) Phase diagram of the gap maximum $\Delta_{\text{max}}$ at zero temperature in terms of sublattice coupling asymmetry $\Omega$ and chemical potential $\mu$. The green, blue and red colors correspond to the spin-triplet $p$-wave, spin-triplet $f$-wave and spin singlet $d$-wave phases, respectively. The hatched areas show where symmetries on both sides of the boundary lead to convergence with the underneath color indicating the symmetry with larger gap amplitude. Here we use the same parameters as indicated for Fig.~\ref{fig:Tc}.
  }%
  \label{fig:T0}
\end{figure*}

Applying the FS average in the nonlinear Eq.~(\ref{eq:gap}), the ensuing gap takes the form
\begin{equation}
    \Delta_{\boldsymbol{k}} = -D(\mu)\langle V^{O(\boldsymbol{k})}_{\boldsymbol{k}\boldsymbol{k^{'}}} \Delta_{\boldsymbol{k}^{'}}\chi(\Delta_{\boldsymbol{k'}}) \rangle_{\boldsymbol{k}^{'},\text{FS}}
\label{eq:gap_T0}
\end{equation}
where 
\begin{equation}
    \chi(\Delta_{\boldsymbol{k'}})=\int_{-\omega_c}^{\omega_c}d\epsilon \frac{\tanh(\sqrt{\epsilon^2+|\Delta_{\boldsymbol{k}^{'}}|^2}/2k_B T)}{2\sqrt{\epsilon^2+|\Delta_{\boldsymbol{k}^{'}}|^2}}.
\end{equation}
Using $\tanh(\sqrt{\epsilon^2+|\Delta_{\boldsymbol{k}^{'}}|^2}/2k_B T)\rightarrow 1$ at $T=0$, we obtain the analytical expression
\begin{equation}
\chi(\Delta_{\boldsymbol{k'}})=\int_0^{\omega_c}\frac{d\epsilon}{\sqrt{\epsilon^2+\Delta_{\boldsymbol{k}^{'}}^2}}=\arsinh\pqty{\frac{\omega_c}{|\Delta_{\boldsymbol{k}^{'}}|}}. 
\label{eq:chi}
\end{equation}
Next, we insert the gap information obtained at $T_c$ in the previous section to construct a trial solution in order to numerically solve the nonlinear Eq.~(\ref{eq:gap_T0}) at zero temperature.
$\Delta_{\text{max}}(0)$ is defined as the largest amplitude of the gap function at zero temperature.
The amplitude of the trial solution is set to $2\omega_c e^{-1/\lambda}$ based on the BCS result for the zero temperature gap amplitude \cite{fossheim2004superconductivity}. 
Fig.~\ref{fig:T0} is the resulting phase diagram for $\Delta_{\text{max}}(0)$, which shows the same superconducting symmetry phases as Fig.~\ref{fig:Tc} at the critical temperature except for small changes of the transition boundaries. 

Close to the phase boundaries we considered both symmetries found at $T=T_c$ for the trial solution and found some regions where both symmetries converge, illustrated by hatched regions. For $0.5 \lesssim \Omega \lesssim 0.9$ a hatched region is not shown at the transition between $p$-wave and $f$-wave. There, $T_c$ and $\Delta_{\text{max}}$ are both from a numerical and an experimental standpoint zero, rendering the determination of a hatched region insignificant. 

The degeneracies in the $p$- and $f$-wave regions found close to $T_c$ are also present at $T=0$. Any linear combination of $p_x$ ($f_x$) and $p_y$ ($f_y$) can solve the zero temperature FS averaged gap equation in the green (blue) region. We conjecture that a more complete momentum resolved zero temperature gap equation may prefer one specific linear combination. The free energy could be used to decide which symmetry is preferred in the hatched regions, and which linear combinations are preferred in the green and blue regions. The free energy depends on $\Delta_{\boldsymbol{k}}$ in the entire 1BZ, while we have solved FS averaged gap equations. With that degree of approximation, we do not believe we have sufficient information to accurately judge the energetically preferred gap symmetry. Since $p_x + ip_y$ gives no nodes on the FS for $E_{\boldsymbol{k}}$ it should give a larger condensation energy than a real combination of $p_x$ and $p_y$. On the other hand, $p_x$ type solutions at zero temperature have a slightly larger amplitude than the $p_x+ip_y$ solutions in this system. We conclude that they are competing orders and name the region $p$-wave. The same goes for linear combinations of $f_x$ and $f_y$ in the $f$-wave region. In the hatched regions, $p$-wave and $f$-wave, or $f$-wave and $d$-wave are competing orders.

There is a possibility that a time-reversal-symmetry-broken state, where the spin triplet gap with zero net spin has $p_x+ip_y$ symmetry, is energetically preferred in the green region of the phase diagram. As explained in Ref.~\cite{Lu2014AFMTSC}, such a fully gapped state has chiral edge states, and can be thought of as a topological superconductor. However, by symmetry arguments, Majorana zero modes in the core of superconducting vortices are not expected \cite{Lu2014AFMTSC}. Hence, it could not find applications in topological quantum computation which involves braiding of Majorana zero modes \cite{TopoQuantumCompRevModPhys, Bernevig2013, TopoSCrevSato}. Topological superconductors with Majorana zero modes in the core of vortices require Cooper pairs made up of spinless fermions or spin polarized electrons \cite{Bernevig2013, TopoSCrevSato}, neither of which are possible at an interface between a NM and a MI with colinear ground states \cite{maeland2022topological}.

References \cite{Schrieffer1989AFMSC, Frenkel1990AFMSC, Ismer2010AFMSC, Lu2014AFMTSC, Rowe2015AFMSC, Romer2016AFMSC} consider the coexistence of antiferromagnetic order and superconductivity in antiferromagnetic metals. The current consensus \cite{Romer2016AFMSC} is that $d$-wave superconductivity is preferred. These studies of bulk systems should involve equal coupling to both spin species of the antiferromagnet. Thus, they should be most comparable to the $\Omega = 1$ case of the AFMI/NM interface in this paper, where we also found $d$-wave pairing.

\section{Conclusion}
In this paper, we have considered superconducting ordering mediated by antiferromagnetic magnons both close to the critical temperature and in the low-temperature regime. We use a BCS weak-coupling approach with an effective magnon-mediated electron-electron interaction. Specifically, we have utilized analytical expressions for effective interactions mediated by such magnons, including properly the effect of Umklapp-processes, which will be important once the Fermi-surface is large enough that any two points on the FS cannot be connected by the momenta of the magnetic Brillouin-zone, available to the magnons. We find that Umklapp processes play an important role in determining the correct superconducting gap-symmetry over large tracts of the $(\mu,\Omega)$-phase diagram. At low to intermediate filling fractions, we find $p$-wave pairing, while for most values of $\Omega$ we find $f$-wave pairing for intermediate to large filling fractions. Close to $\Omega=1$, we find $d$-wave pairing. The results obtained at $T=T_c$ correspond well to the results obtained at $T=0$. Hence, the symmetries found for the superconducting gap-function are stable to variation in temperature. Our results at $T=T_c$ obtained within a BCS weak-coupling approach are in very good agreement with previous results found using a much more elaborate strong-coupling Eliashberg approach.

\begin{acknowledgments} 
We acknowledge funding from the Research Council of Norway (RCN) through its Centres of Excellence funding scheme Project No.~262633, ``QuSpin," and RCN Project No.~323766, ``Equilibrium and out-of-equilibrium quantum phenomena in superconducting hybrids with antiferromagnets and topological insulators."
\end{acknowledgments}

\appendix
\section{Holstein-Primakoff and Fourier transformations} \label{app:HPandFT}
We introduce the Holstein-Primakoff transformation for the two sublattice spin operators in the AFMI
\begin{align}
    S_{i+}^A &= S_{ix}^A+iS_{iy}^A= \sqrt{2s-a_i^\dag a_i}a_i \approx \sqrt{2s}a_i,\\ 
    S_{i-}^A &= S_{ix}^A-iS_{iy}^A= a_i^\dag\sqrt{2s-a_i^\dag a_i} \approx \sqrt{2s}a_i^\dag,\\ 
    S_{iz}^A &= s -a_i^\dag a_i,
\end{align}
\begin{align}
     S_{j+}^B &= S_{jx}^B+iS_{jy}^B= b_j^\dag\sqrt{2s-b_j^\dag b_j} \approx \sqrt{2s}b_j^\dag,\\ 
     S_{j-}^B &= S_{jx}^B-iS_{jy}^B= \sqrt{2s-b_j^\dag b_j}b_j \approx \sqrt{2s}b_j,\\ 
    S_{jz}^B &= -s +b_j^\dag b_j,  
\end{align}
where $a_i$ and $b_j$ are the two sublattice magnon operators and $s$ is the spin quantum number associated with the lattice site spins. Next, we perform the Fourier transformations of the magnon operators 
\begin{equation}
    a_i = \frac{1}{\sqrt{N_A}} \sum_{\boldsymbol{q} \in \Diamond} a_{\boldsymbol{q}}e^{-i\boldsymbol{q}\cdot \boldsymbol{r}_i},\quad b_i = \frac{1}{\sqrt{N_B}} \sum_{\boldsymbol{q} \in \Diamond} b_{\boldsymbol{q}}e^{-i\boldsymbol{q}\cdot \boldsymbol{r}_i},
\end{equation}
in which $N_A = N_B = N/2$ and $N$ is the number of lattice sites at the AFMI/NM interface. $\Diamond$ is used to mark the sum over the RBZ. 

Inserting these transformations directly in $H_{\text{int}}$ yields some additional terms that have been ignored compared to Eqs.~\eqref{eq:HintA} and \eqref{eq:HintB}. First, terms involving two magnon operators are assumed negligible compared to the one-magnon processes. 
Second, terms involving only electron operators, akin to Zeeman terms, are expected to have little influence on our results regarding the superconducting state when the spin splitting is small compared to the electron bandwidth, i.e., $\bar{J}\ll t$ \cite{erlandsen2019enhancement}. They are only nonzero for $\Omega < 1$, and can be eliminated by considering an AFMI/NM/AFMI trilayer \cite{thingstad2021eliashberg}, or a compensating magnetic field to ensure that the superconducting state is preferred over the normal state, cf.~the Chandrasekhar-Clogston limit. For the bilayer case and $\Omega < 1$, this implies that one can observe the {\it appearance} of superconductivity when applying an external magnetic field. 

\section{Derivation of effective interaction} \label{app:HPair}
Here we derive the effective interaction mediated by the antiferromagnetic magnons. We use a Schrieffer-Wolff transformation to obtain an effective magnon-mediated interaction to second order in the coupling constant $\bar{J}$ between the AFMI and the electron spins in the NM. This is effectuated by a canonical transformation $H^{'}= e^{-\eta S} H e^{\eta S}$, such that the pairing interaction to second order in $V \equiv -2\bar{J}\sqrt{s}/\sqrt{N}$ becomes  
\begin{equation}
    H_\text{pair} = \sum_{LL^{'}} H_\text{pair}^{(L,L^{'})} = \sum_{LL^{'}}\frac{1}{2}[\eta H_1^{(L)}, \eta S^{(L^{'})}],
\label{eq:Hpair}
\end{equation}
where $L\in \{A,B\}$ and $\eta S^{(L)}$ satisfies
\begin{equation}
    \eta H_1^{(L)} + [H_0,\eta S^{(L)}]=0.
    \label{eq:S}
\end{equation}
\begin{widetext}
Choosing the ansatze
\begin{align}
    \eta S^{(A)} = \Omega V \sum_{\boldsymbol{k}\boldsymbol{q}}&[(x_{\boldsymbol{k},\boldsymbol{q}}u_{\boldsymbol{q}}\alpha_{\boldsymbol{q}} + y_{\boldsymbol{k},\boldsymbol{q}}v_{\boldsymbol{q}}\beta_{-\boldsymbol{q}}^\dag)c_{\boldsymbol{k}+\boldsymbol{q},\downarrow}^\dag c_{\boldsymbol{k}\uparrow} + (x_{\boldsymbol{k},\boldsymbol{q},\boldsymbol{G}}u_{\boldsymbol{q}}\alpha_{\boldsymbol{q}} + y_{\boldsymbol{k},\boldsymbol{q},\boldsymbol{G}}v_{\boldsymbol{q}}\beta_{-\boldsymbol{q}}^\dag)c_{\boldsymbol{k}+\boldsymbol{q}+\boldsymbol{G},\downarrow}^\dag c_{\boldsymbol{k}\uparrow}\notag\\
    &+ (z_{\boldsymbol{k},\boldsymbol{q}}u_{\boldsymbol{q}}\alpha_{-\boldsymbol{q}}^\dag + w_{\boldsymbol{k},\boldsymbol{q}}v_{\boldsymbol{q}}\beta_{\boldsymbol{q}})c_{\boldsymbol{k}+\boldsymbol{q},\uparrow}^\dag c_{\boldsymbol{k}\downarrow} + (z_{\boldsymbol{k},\boldsymbol{q},\boldsymbol{G}}u_{\boldsymbol{q}}\alpha_{-\boldsymbol{q}}^\dag + w_{\boldsymbol{k},\boldsymbol{q},\boldsymbol{G}}v_{\boldsymbol{q}}\beta_{\boldsymbol{q}})c_{\boldsymbol{k}+\boldsymbol{q}+\boldsymbol{G},\uparrow}^\dag c_{\boldsymbol{k}\downarrow}],
\end{align}
\begin{align}
   \eta S^{(B)} = V \sum_{\boldsymbol{k}\boldsymbol{q}}&[(w_{\boldsymbol{k},\boldsymbol{q}}u_{\boldsymbol{q}}\beta_{\boldsymbol{q}} + z_{\boldsymbol{k},\boldsymbol{q}}v_{\boldsymbol{q}}\alpha_{-\boldsymbol{q}}^\dag)c_{\boldsymbol{k}+\boldsymbol{q},\uparrow}^\dag c_{\boldsymbol{k}\downarrow} - (w_{\boldsymbol{k},\boldsymbol{q},\boldsymbol{G}}u_{\boldsymbol{q}}\beta_{\boldsymbol{q}} + z_{\boldsymbol{k},\boldsymbol{q},\boldsymbol{G}}v_{\boldsymbol{q}}\alpha_{-\boldsymbol{q}}^\dag)c_{\boldsymbol{k}+\boldsymbol{q}+\boldsymbol{G},\uparrow}^\dag c_{\boldsymbol{k}\downarrow}\notag\\
    &+ (y_{\boldsymbol{k},\boldsymbol{q}}u_{\boldsymbol{q}}\beta_{-\boldsymbol{q}}^\dag + x_{\boldsymbol{k},\boldsymbol{q}}v_{\boldsymbol{q}}\alpha_{\boldsymbol{q}})c_{\boldsymbol{k}+\boldsymbol{q},\downarrow}^\dag c_{\boldsymbol{k}\uparrow} - (y_{\boldsymbol{k},\boldsymbol{q},\boldsymbol{G}}u_{\boldsymbol{q}}\beta_{-\boldsymbol{q}}^\dag + x_{\boldsymbol{k},\boldsymbol{q},\boldsymbol{G}}v_{\boldsymbol{q}}\alpha_{\boldsymbol{q}})c_{\boldsymbol{k}+\boldsymbol{q}+\boldsymbol{G},\downarrow}^\dag c_{\boldsymbol{k}\uparrow}], 
\end{align}
and inserting them into Eq.~(\ref{eq:S}), the coefficients in the ansatze can be solved as 
\begin{align}
x_{\boldsymbol{k},\boldsymbol{q}} = w_{\boldsymbol{k},\boldsymbol{q}} &= \frac{1}{\epsilon_{\boldsymbol{k}} - \epsilon_{\boldsymbol{k}+\boldsymbol{q}} + \omega_{\boldsymbol{q}}},
\label{xkq} 
&
x_{\boldsymbol{k},\boldsymbol{q},\boldsymbol{G}} = w_{\boldsymbol{k},\boldsymbol{q},\boldsymbol{G}} &= \frac{1}{\epsilon_{\boldsymbol{k}} - \epsilon_{\boldsymbol{k}+\boldsymbol{q}+\boldsymbol{G}} + \omega_{\boldsymbol{q}}}, 
 \\
y_{\boldsymbol{k},\boldsymbol{q}} = z_{\boldsymbol{k},\boldsymbol{q}} &= \frac{1}{\epsilon_{\boldsymbol{k}} - \epsilon_{\boldsymbol{k}+\boldsymbol{q}} - \omega_{\boldsymbol{q}}},
%
&
y_{\boldsymbol{k},\boldsymbol{q},\boldsymbol{G}} = z_{\boldsymbol{k},\boldsymbol{q},\boldsymbol{G}} &= \frac{1}{\epsilon_{\boldsymbol{k}} - \epsilon_{\boldsymbol{k}+\boldsymbol{q}+\boldsymbol{G}} - \omega_{\boldsymbol{q}}}.
\label{ykqg}
\end{align}

Given $\eta S^{(L)}$, the pairing effective electron-electron interactions given by Eq.~(\ref{eq:Hpair}) are calculated as
\begin{align}
    H_{\text{pair}}^{(A,A)} = \frac{1}{2} \Omega^2 V^2 \sum_{\boldsymbol{k}\boldsymbol{k}^{'}\boldsymbol{q}}\{ &[u_{\boldsymbol{q}}^2 (y_{\boldsymbol{k}^{'},-\boldsymbol{q}} - x_{\boldsymbol{k},\boldsymbol{q}}) + v_{\boldsymbol{q}}^2 (y_{\boldsymbol{k},\boldsymbol{q}} - x_{\boldsymbol{k}^{'},-\boldsymbol{q}})]c_{\boldsymbol{k}+\boldsymbol{q},\downarrow}^\dag c_{\boldsymbol{k}\uparrow} c_{\boldsymbol{k}^{'}-\boldsymbol{q},\uparrow}^\dag c_{\boldsymbol{k}^{'}\downarrow} \notag\\
    &+ [u_{\boldsymbol{q}}^2 (y_{\boldsymbol{k}^{'},-\boldsymbol{q},\boldsymbol{G}} - x_{\boldsymbol{k},\boldsymbol{q},\boldsymbol{G}}) + v_{\boldsymbol{q}}^2 (y_{\boldsymbol{k},\boldsymbol{q},\boldsymbol{G}} - x_{\boldsymbol{k}^{'},-\boldsymbol{q},\boldsymbol{G}})]c_{\boldsymbol{k}+\boldsymbol{q}+\boldsymbol{G},\downarrow}^\dag c_{\boldsymbol{k}\uparrow} c_{\boldsymbol{k}^{'}-\boldsymbol{q}+\boldsymbol{G},\uparrow}^\dag c_{\boldsymbol{k}^{'}\downarrow} \},
\end{align}
\begin{align}
   H_{\text{pair}}^{(B,B)} = \frac{1}{2}  V^2 \sum_{\boldsymbol{k}\boldsymbol{k}^{'}\boldsymbol{q}}\{ &[u_{\boldsymbol{q}}^2 (y_{\boldsymbol{k},\boldsymbol{q}} - x_{\boldsymbol{k}^{'},-\boldsymbol{q}}) + v_{\boldsymbol{q}}^2 (y_{\boldsymbol{k}^{'},-\boldsymbol{q}} - x_{\boldsymbol{k},\boldsymbol{q}})]c_{\boldsymbol{k}+\boldsymbol{q},\downarrow}^\dag c_{\boldsymbol{k}\uparrow} c_{\boldsymbol{k}^{'}-\boldsymbol{q},\uparrow}^\dag c_{\boldsymbol{k}^{'}\downarrow} \notag\\
    &+ [u_{\boldsymbol{q}}^2 (y_{\boldsymbol{k},\boldsymbol{q},\boldsymbol{G}} - x_{\boldsymbol{k}^{'},-\boldsymbol{q},\boldsymbol{G}}) + v_{\boldsymbol{q}}^2 (y_{\boldsymbol{k}^{'},-\boldsymbol{q},\boldsymbol{G}} - x_{\boldsymbol{k},\boldsymbol{q},\boldsymbol{G}})]c_{\boldsymbol{k}+\boldsymbol{q}+\boldsymbol{G},\downarrow}^\dag c_{\boldsymbol{k}\uparrow} c_{\boldsymbol{k}^{'}-\boldsymbol{q}+\boldsymbol{G},\uparrow}^\dag c_{\boldsymbol{k}^{'}\downarrow} \}, 
\end{align}
\begin{align}
    H_{\text{pair}}^{(A,B)} + H_{\text{pair}}^{(B,A)} = \Omega V^2 \sum_{\boldsymbol{k}\boldsymbol{k}^{'}\boldsymbol{q}} &[u_{\boldsymbol{q}}v_{\boldsymbol{q}}(y_{\boldsymbol{k},\boldsymbol{q}} + y_{\boldsymbol{k}^{'},-\boldsymbol{q}} - x_{\boldsymbol{k},\boldsymbol{q}} - x_{\boldsymbol{k}^{'},-\boldsymbol{q}})c_{\boldsymbol{k}+\boldsymbol{q},\downarrow}^\dag c_{\boldsymbol{k}\uparrow} c_{\boldsymbol{k}^{'}-\boldsymbol{q},\uparrow}^\dag c_{\boldsymbol{k}^{'}\downarrow}\notag\\
    &-u_{\boldsymbol{q}}v_{\boldsymbol{q}}(y_{\boldsymbol{k},\boldsymbol{q},\boldsymbol{G}} + y_{\boldsymbol{k}^{'},-\boldsymbol{q},\boldsymbol{G}} - x_{\boldsymbol{k},\boldsymbol{q},\boldsymbol{G}} - x_{\boldsymbol{k}^{'},-\boldsymbol{q},\boldsymbol{G}})c_{\boldsymbol{k}+\boldsymbol{q}+\boldsymbol{G},\downarrow}^\dag c_{\boldsymbol{k}\uparrow} c_{\boldsymbol{k}^{'}-\boldsymbol{q}+\boldsymbol{G},\uparrow}^\dag c_{\boldsymbol{k}^{'}\downarrow}].
\end{align}
\end{widetext}
We next assume that only electrons with opposite momenta interact. Inserting the expressions in Eqs.~(\ref{xkq}-\ref{ykqg}), we then obtain the effective electron-electron interaction for scattering electron-pairs from $(\boldsymbol{k}',-\boldsymbol{k}')$ to $({\boldsymbol{k}},-{\boldsymbol{k}})$, given in Eq.~\eqref{eq:HpairMain}. 

\section{Density of states in the NM}\label{app:DOS}
Following Refs.~\cite{maeland2021electron, maeland2022topological}, the density of states per spin in the NM can be calculated as 
\begin{equation}
    D(\epsilon) = \sum_{\boldsymbol{k}\in \Box} \delta(\epsilon - \epsilon_{\boldsymbol{k}}) = \frac{N}{A_\Box}\int_{-\pi}^{\pi}d\theta \int_{0}^{c(\theta)} dk k \delta(\epsilon-\epsilon_{k,\theta}),
\end{equation}
in which $\theta = \operatorname{atan2}(k_y/k_x)$ with $k_x=k\cos\theta$ and $k_y = k\sin\theta$. $A_\Box=4\pi^2$ denotes the area of the full Brillouin zone $\Box$ and the upper cutoff $c(\theta) = \pi / \text{max} \{|\cos\theta|,|\sin\theta|\}$ confines the integral to $\boldsymbol{k}\in \Box$.

Given a function $f(k)$ with roots $k_i$ and $f^{\prime}(k_i)\neq 0$, $\delta[f(k)]=\sum_i \delta(k-k_i)/|f^{\prime}(k_i)|$. Here we have $f(k) = \epsilon + 2t[\cos(k\cos\theta)+\cos(k\sin\theta)]$ based on Eq.~(\ref{eq:FS}) with $\mu = 0$. Consequently, we have
\begin{equation}
    D(\epsilon) = \frac{N}{A_\Box}\int_{-\pi}^{\pi}d\theta \sum_i \frac{k_i(\theta)}{|f^{\prime}[k_i(\theta)]|}.
\label{eq:DOS(epsilon)}
\end{equation}
Since $\epsilon$ and $\mu$ enter the above equations in the same way, we can use Eq.~(\ref{eq:DOS(epsilon)}) from setting $\mu=0$ to obtain the density of states on the FS ($\epsilon=0$) at nonzero $\mu$ as $D(\epsilon = \mu) \equiv D(\mu)$.

\bibliography{main.bbl}
\end{document}